\documentclass[a4paper,12pt, epsfig]{article}
\usepackage{epsfig}
\usepackage{amssymb,latexsym}
\usepackage{amsfonts}
\usepackage{amsmath}

\newskip\humongous \humongous=0pt plus 1000pt minus 1000pt

\newif\ifdtup

\catcode`\@=11

\@addtoreset{equation}{section}
\def\theequation{\thesection.\arabic{equation}}

\def\@normalsize{\@setsize\normalsize{15pt}\xiipt\@xiipt
\abovedisplayskip 14pt plus3pt minus3pt%
\belowdisplayskip \abovedisplayskip
\abovedisplayshortskip \z@ plus3pt%
\belowdisplayshortskip 7pt plus3.5pt minus0pt}

\def\small{\@setsize\small{13.6pt}\xipt\@xipt
\abovedisplayskip 13pt plus3pt minus3pt%
\belowdisplayskip \abovedisplayskip
\abovedisplayshortskip \z@ plus3pt%
\belowdisplayshortskip 7pt plus3.5pt minus0pt
\def\@listi{\parsep 4.5pt plus 2pt minus 1pt
      \itemsep \parsep
      \topsep 9pt plus 3pt minus 3pt}}

\relax

\catcode`@=12

\catcode`\@=11

\def\section{\@startsection{section}{1}{\z@}{3.5ex plus 1ex minus
    .2ex}{2.3ex plus .2ex}{\large\bf}}

\def\thesection{\arabic{section}}
\def\thesubsection{\arabic{section}.\arabic{subsection}}

\def\appendix{\setcounter{section}{0}
  \def\thesection{Appendix \Alph{section}}
  \def\thesubsection{\Alph{section}.\arabic{subsection}}
  \def\theequation{\Alph{section}.\arabic{equation}}}

\def\SymBoxes#1#2#3#4{\newdimen\un@t \un@t#3%
\raisebox{#1}{\rule{#2\un@t}{#4}\hskip-#2\un@t
\@tempdimb\un@t \advance\@tempdimb by-#4\@tempcntb#2\relax%
\@whilenum{\@tempcntb>0}\do{
\rule{#4}{\un@t}\hskip\@tempdimb \advance\@tempcntb by\m@ne}%
\hskip-#2\un@t \rule[\un@t]{#2\un@t}{#4}%
\rule[\un@t]{#4}{#4}\hskip-#4
\rule{#4}{\un@t}}\hskip-#4}                

\begin{document}


\newcommand{\dd}{\textrm{d}}

\newcommand{\beq}{\begin{equation}}
\newcommand{\eeq}{\end{equation}}
\newcommand{\bea}{\begin{eqnarray}}
\newcommand{\eea}{\end{eqnarray}}
\newcommand{\beas}{\begin{eqnarray*}}
\newcommand{\eeas}{\end{eqnarray*}}
\newcommand{\defi}{\stackrel{\rm def}{=}}
\newcommand{\non}{\nonumber}
\newcommand{\bquo}{\begin{quote}}
\newcommand{\enqu}{\end{quote}}
\renewcommand{\(}{\begin{equation}}
\renewcommand{\)}{\end{equation}}
\def\de{\partial}
\def\Om{\ensuremath{\Omega}}
\def\Tr{ \hbox{\rm Tr}}
\def\rc{ \hbox{$r_{\rm c}$}}
\def\H{ \hbox{\rm H}}
\def\HE{ \hbox{$\rm H^{even}$}}
\def\HO{ \hbox{$\rm H^{odd}$}}
\def\HEO{ \hbox{$\rm H^{even/odd}$}}
\def\HOE{ \hbox{$\rm H^{odd/even}$}}
\def\HHO{ \hbox{$\rm H_H^{odd}$}}
\def\HHEO{ \hbox{$\rm H_H^{even/odd}$}}
\def\HHOE{ \hbox{$\rm H_H^{odd/even}$}}
\def\K{ \hbox{\rm K}}
\def\Im{ \hbox{\rm Im}}
\def\Ker{ \hbox{\rm Ker}}
\def\const{\hbox {\rm const.}}
\def\o{\over}
\def\im{\hbox{\rm Im}}
\def\re{\hbox{\rm Re}}
\def\bra{\langle}\def\ket{\rangle}
\def\Arg{\hbox {\rm Arg}}
\def\exo{\hbox {\rm exp}}
\def\diag{\hbox{\rm diag}}
\def\longvert{{\rule[-2mm]{0.1mm}{7mm}}\,}
\def\a{\alpha}
\def\b{\beta}
\def\e{\epsilon}
\def\l{\lambda}
\def\ol{{\overline{\lambda}}}
\def\ochi{{\overline{\chi}}}
\def\th{\theta}
\def\s{\sigma}
\def\oth{\overline{\theta}}
\def\ad{{\dot{\alpha}}}
\def\bd{{\dot{\beta}}}
\def\oD{\overline{D}}
\def\opsi{\overline{\psi}}
\def\dag{{}^{\dagger}}
\def\tq{{\widetilde q}}
\def\L{{\mathcal{L}}}
\def\p{{}^{\prime}}
\def\W{W}
\def\N{{\cal N}}
\def\hsp{,\hspace{.7cm}}
\def\bo{\ensuremath{\hat{b}_1}}
\def\bfo{\ensuremath{\hat{b}_4}}
\def\co{\ensuremath{\hat{c}_1}}
\def\cfo{\ensuremath{\hat{c}_4}}
\newcommand{\C}{\ensuremath{\mathbb C}}
\newcommand{\Z}{\ensuremath{\mathbb Z}}
\newcommand{\R}{\ensuremath{\mathbb R}}
\newcommand{\rp}{\ensuremath{\mathbb {RP}}}
\newcommand{\cp}{\ensuremath{\mathbb {CP}}}
\newcommand{\vac}{\ensuremath{|0\rangle}}
\newcommand{\vact}{\ensuremath{|00\rangle}                    }
\newcommand{\oc}{\ensuremath{\overline{c}}}

\newcommand{\Vol}{\textrm{Vol}}

\newcommand{\half}{\frac{1}{2}}

\def\changed#1{{\bf #1}}

\begin{titlepage}
\begin{flushright}
IFUP-TH/2010-27
\end{flushright}
\bigskip
\def\thefootnote{\fnsymbol{footnote}}

\begin{center}
{\large {\bf
The Surface Layers Dual to Hydrodynamic Boundaries
  } }
\end{center}

\bigskip
\begin{center}
{\large  Jarah Evslin\footnote{\texttt{jarah@df.unipi.it}} and Giovanni Ricco\footnote{\texttt{giovanni.ricco@gmail.com}}}
\end{center}

\renewcommand{\thefootnote}{\arabic{footnote}}

\begin{center}
\vspace{0em}
{\em  { INFN Sezione di Pisa, Largo Pontecorvo 3, Ed. C, 56127 Pisa, Italy\\and\\
Dipartimento di Fisica, Univ di Pisa, Largo Pontecorvo 3, Ed. C, 56127 Pisa, Italy\\
\vskip .4cm}}

\end{center}

\vspace{1.1cm}

\noindent
\begin{center} {\bf Abstract} \end{center}

\noindent
The AdS/hydrodynamics correspondence provides a 1-1 map between large wavelength features of AdS black branes and conformal fluid flows.  In this note we consider boundaries between nonrelativistic flows, applying the usual boundary conditions for viscous fluids.  We find that a naive application of the correspondence to these boundaries yields a surface layer in the gravity theory whose stress tensor is not equal to that given by the Israel matching conditions.  In particular, while neither stress tensor satisfies the null energy condition and both have nonvanishing momentum, only Israel's tensor has stress.  The disagreement arises entirely from corrections to the metric due to multiple derivatives of the flow velocity, which violate Israel's finiteness assumption in the thin wall limit.  


\vfill

\begin{flushleft}
{\today}
\end{flushleft}
\end{titlepage}

\hfill{}


\setcounter{footnote}{0}

\section{Introduction}
It has long been known that the dynamics of a $p$-dimensional gravitational theory is captured by quantities on $(p-1)$-dimensional hypersurfaces \cite{EIH}.  It was argued by Damour \cite{Damour}, based on an analogy by Hartle and Hawking \cite{HH}, that in the case of certain black hole solutions these surface quantities describe the flow of a viscous $(p-1)$-dimensional fluid.  Perhaps the most concrete realization of this idea is the one to one map between large wavelength features of asymptotically AdS$_p$ black brane solutions and $(p-1)$-dimensional conformal fluid flows presented recently in Refs.~\cite{BHMR} and \cite{arbdim}.  

This AdS/hydrodynamics correspondence provides an explicit black brane solution for every history of a particular conformal fluid, so long as the fluid variables are constant over distances large compared with the inverse temperature.  For example progress towards gravity duals of shock waves and vortices has appeared in Refs.~\cite{shock} and \cite{vortices}.  In particular, there must be gravity duals to turbulent flows.  Turbulence is generic in fluid flows under a wide range of conditions.  The dual of these fluid conditions then provides some condition on a gravity solution under which it to generically decays into a turbulent configuration.  An example of such a situation was presented in Ref.~\cite{forced}.  It would of course be interesting to characterize the gravity duals of turbulent flows, and of the conditions under which turbulence may be expected.  In hydrodynamics, even the most basic scaling laws are altered by turbulence.  If gravitational solutions near, for example, spacelike singularities (where indeed chaotic evolution is expected \cite{BKL}) or certain event horizons do generically decay to turbulent solutions, it would be difficult to overstate the potential consequences for, for example, the horizon problem.

Perhaps the best understood turbulence is steady state turbulence, in which energy is injected into a system at the same rate at which it dissipates.  Richardson's cascade model \cite{Rich} of steady (3+1)-dimensional turbulence is as follows.  Energy is injected into a system at large characteristic distance scales, for example, a lake warms the air.  This creates large vortices, which decay into smaller vortices.  Thus the energy flows to smaller distance scales.  At sufficiently small distance scales, higher order derivative terms in the equations of motion become relevant, such as viscosity terms.  These lead to dissipation of the energy in sufficiently small vortices.  Thus energy cascades from the long length scale in which it is introduced, down to the dissipation scale.

To realize steady state turbulence, one needs to inject energy into a system.  There are two principal ways to do this.  First, one may deform the fluid via external perturbations.  Second, one may apply boundary conditions, for example one may consider fluid flow in a pipe or wind tunnel.  The first approach was applied to the AdS/hydrodynamics correspondence in Ref.~\cite{forced}, where it was argued that a laminar fluid flow and the dual gravity solution decay to turbulent configurations.  This approach has the disadvantage that solutions are quite complicated, due to the necessarily inhomogeneous forcing and to the geometric implementation of the forcing on the gravity side.  

In this note we will take a preliminary step towards a realization of the second approach to creating steady state turbulence, we will investigate boundary conditions in the AdS/hydrodynamics correspondence.  For simplicity, we will consider nonrelativistic, incompressible flows.  Consider the surface which separates a solid object from such a fluid.  The normal velocity of the fluid into the solid must vanish.  If furthermore the fluid is viscous, as fluids in the AdS/hydrodynamics correspondence are \cite{Damour}, then the tangential relative velocity of the fluid must also vanish.

What does this correspond to on the gravity side?  The answer to this question is not necessarily unique, one may define a dual and then attempt to understand its dynamics.  One interesting case, which is already sufficient to generate turbulence, is a solid which is a thin, infinite sheet with a stationary fluid on the left side and a moving fluid on the right.  In this case a natural choice would be to consider the gravity duals of both fluids and then to attempt to glue them together.  Equivalently one may choose to think of the entirety of the left side as a solid wall, filling the left half of spacetime, and a liquid filling the right half.  The wall is stationary and so one chooses the dual to be a stationary black brane in half of AdS.  Whatever one chooses to think, the logic is that one imposes that the left half of the gravity dual be a static black brane in AdS, and that the right side be the gravity dual given by the prescription of Ref.~\cite{BHMR}.

So how does one glue these two vacuum gravity solutions together?  Clearly there are many inequivalent choices.  One possibility is to simply attach them and then use the Israel matching conditions \cite{Israel} to determine the stress tensor on the surface layer that separates the two sides.  This is equivalent to letting the gravitational solution continuously interpolate between the two solutions over a finite distance $d$ and then taking the limit as this distance tends to zero.  While there are many ways of performing this interpolation, so long as the extrinsic curvature is kept finite, they all lead to the same stress tensor as the interpolation distance $d\rightarrow 0$.

Another possibility is to let the fluid configuration continuously interpolate between the two solutions, and then take the dual using the prescription of Ref.~\cite{BHMR}.  As the fluid is not a solution of the Navier-Stokes equation in this region, the dual will not be a solution of the vacuum Einstein equations in this region.  Instead it will solve Einstein's equations with a nonvanishing stress tensor supported on a surface layer.  The ultralocality of the duality map implies that the vacuum Einstein equations will however be solved away from the surface layer.  In this case, one cannot take the interpolation distance $d$ to zero, because the dual is not defined when derivatives are large with respect to the inverse of the temperature $T$.  Thus the minimum size of $d$ will be of order $1/T$.  Again there are many inequivalent ways of performing the interpolation.  But we will see that, at least for the quantities at we are able to calculate, when $d$ is large with respect to $1/T$, the difference between these prescriptions is suppressed by powers of $dT$ and so, like Israel's method, there is a single answer.

The perhaps surprising result is that the two methods yield bulk stress tensors which differ by a finite amount.  They did not need to agree, indeed one is derived at small $d$ and the other for large $d$.  The reason that they disagree is as follows.  The construction of the metric from the fluid flow proceeds order by order in the derivatives of the fluid's velocity.  The boundary conditions imply that the velocity of the fluid is the same on both sides of the wall, however the first derivatives differ.  Therefore, whatever regularization scheme one uses on the fluid side, the second derivative of the velocity diverges at small $d$.  This means that the metric corrections derived using the map of \cite{BHMR} will diverge at small $d$, invalidating the finiteness assumption in Israel's derivation.  In fact, we will see that the disagreement between the two calculations of the stress tensor differ only in these higher derivative terms.  Of course the fluid map is not defined at small $d$, as it yields a divergent series, and so no divergences appear within the range of validity of either approach.

We will begin in Sec.~\ref{flowsez} by describing the flow of interest.  The velocity will be kept sufficiently arbitrary to allow a general interpolation between the flows on the two sides of the wall, and in Sec.~\ref{gravsez} the naive gravity dual will be calculated using the prescription of \cite{BHMR}.  We will see that those higher order derivative corrections which we calculate are indeed suppressed by factors of $dT$.  Then in Sec.~\ref{tsez} we will calculate the bulk stress tensor of the interpolation between the two gravity solutions.  First it will be calculated for the interpolation dual to a continuously interpolating fluid flow.  It will be seen that contributions from the second derivative of the velocity are $d$-independent, while higher order contributions are suppressed by powers of $dT$.  Thus the result is independent of the interpolation scheme when $d$ is sufficiently large.  The stress tensor will then be calculated directly from the Israel matching conditions on the two solutions of the vacuum Einstein equations.  It will be seen that the two stress tensors agree up to terms corresponding to a divergence in the extrinsic curvature at small $d$, and that only the second stress tensor contains a nonvanishing stress.


\section{The Flow} \label{flowsez}

\subsection{The ansatz}

We will consider a hydrodynamic flow in 4-dimensional Minkowski space, using a $(-,+,+,+)$ metric. 
To highlight the essential features of the boundary condition, we will consider the simplest possible flow.  The liquid will only move in the $y$ direction, with a velocity $v=v(x)$ that only depends on the coordinate $x$.  The velocity will be taken to be small, and we will drop all terms which are quadratic in $v$.  In fact, as described in Refs.~\cite{incomp,vortices} we will work in the nonrelativistic, incompressible limit.  More precisely, we will show that our flow satisfies both the full relativistic equations of motion at order $\mathcal{O}(v)$ and also the incompressible Navier-Stokes equation.

We will set $c=1$.  The conformal fluid which is dual to Einstein gravity with a negative cosmological constant is very particular.  Being conformal, all of its transport coefficients may be expressed in terms of a single dimensionful quantity, such as the temperature $T$, and certain constants which may be calculated from the gravity dual.  In the case at hand for example the shear viscosity $\eta$, pressure $p$ and density $\rho$ have been found in Ref.~\cite{BHMR}
\beq
\eta=\frac{\pi^2}{16G_N}T^3\hsp
p=\frac{\pi^3}{16G_N}T^4\hsp
\rho=\frac{3\pi^3}{16G_N}T^4 \label{rapporti}
\eeq
where $G_N$ is the dual Newton's constant.  

\begin{figure}
\begin{center}
\includegraphics[scale=.58]{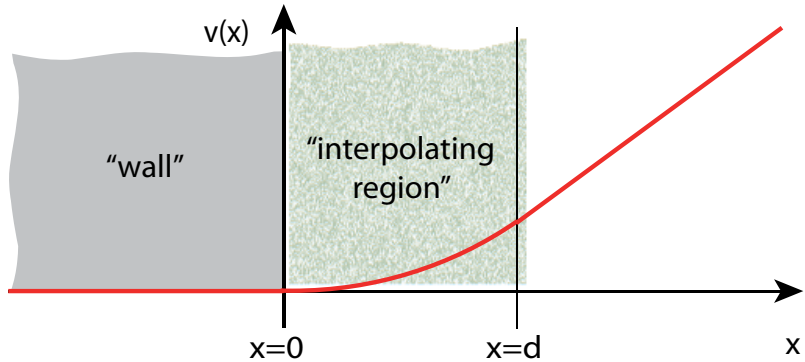}
\caption{The fluid velocity $v$ is in the $y$ direction, and it depends on the $x$ coordinate.  On the left the fluid is stationary, on the right the fluid velocity is linear.  These two regions solve the fluid equations of motion at linear order in $v$.  There is an interpolating region of width $d$, which must be larger than the inverse temperature, in which $v$ does not satisfy the equations of motion.  $v$ and its first derivative $v\p$ are continuous at $x=0$ and $x=d$.}
\label{vy}
\end{center}
\end{figure}

The relativistic velocity 4-vector $u$ is, to linear order in $v$, simply 
\begin{equation}
u_\mu = (\frac{1}{\sqrt{1-v^2}}, 0, \frac{v}{\sqrt{1-v^2}}, 0) \sim (1,0,v,0). \label{u}
\end{equation}
We will be interested in the fluid velocity in three regions, as illustrated in Fig.~\ref{vy}.  First, on the left, where $v=0$.  Second, we will be interested in the velocity on the right, where $v$ will be linear in $x$.  We will show momentarily that this is a solution to the hydrodynamic equations of motion and so will be dual to a vacuum solution of Einstein's equations.  Finally, we will be interested in an interpolating region where $v$ will be arbitrary and we will not impose the equations of motion, therefore the dual metric will not solve the vacuum Einstein equations but, like any metric, will solve Einstein's equations with some stress tensor.

Clearly the left, $v=0$, satisfies the fluid equations of motion.  We will now verify that the region on the right satisfies the relativistic equations of motion, which are simply the conservation of the stress tensor
\beq
0=\partial_\mu T^{\mu\nu}. \label{edm}
\eeq
In accordance with the usual fluid approximation \cite{LL}, we will work at large enough distance scales that only the velocity $v$ and its first derivative $v\p$ need be considered in the stress tensor.  This approximation in general is problematic, leading for example to superluminal propagation \cite{ac}.  However, as we will be interested in velocities well below the speed of light, no problems will arise.  Later, when we will consider the interpolating region, where the second derivative may be large, we will make no such approximation.  We will consider the bulk stress tensor to higher order, calculating all terms up to two derivatives and several terms up to three or four derivatives to check that they are subdominant.  However we do not impose that the interpolating region satisfies the equations of motion, indeed that would lead to a vanishing bulk stress tensor.

\subsection{Relativistic and nonrelativistic equations of motion}

Dropping all higher derivatives of the velocity and using the fact that the fluid is conformal to eliminate the bulk viscosity and replace $\rho$ with $3p$, the hydrodynamic stress tensor is
\beq
T^{\mu\nu}=p(\eta^{\mu\nu}+4u^\mu u^\nu)-2\eta\sigma^{\mu\nu} \label{Tfluido}
\eeq
where the shear strain rate $\sigma^{\mu \nu}$ is defined as
\begin{equation}
\sigma^{\mu \nu} = P^{\mu \alpha} P^{\nu \beta} \partial_{(\alpha} u_{\beta)} - \frac{1}{3}\partial_\lambda u^\lambda P^{\mu \nu}. \label{sigdef} 
\end{equation}
Here parenthesis denote symmetrization with a factor of one half and $P^{\mu\nu}$ is a projector onto the spacelike directions in the reference frame of the fluid
\begin{equation}
P_{\mu \nu} = \eta_{\mu \nu} + u_\mu u_\nu 
= \left(
\begin{array}{cccc}
 \frac{v^2}{1- v^2} & 0 & \frac{v}{1- v^2} & 0 \\
 0 & 1  & 0 & 0 \\
 \frac{v}{1- v^2} & 0 & \frac{1}{1- v^2} & 0\\
 0 & 0 & 0 & 1
\end{array}
\right). \label{proj}
\end{equation}

Substituting the velocity ansatz (\ref{u}) into the definition (\ref{sigdef}) one easily finds the shear strain at linear order in $v$
\begin{equation}
\sigma^{\mu \nu} \simeq \left(
\begin{array}{cccc}
 0 & 0 & 0 & 0 \\
 0 & 0  & \frac{1}{2} v' & 0 \\
 0 & \frac{1}{2} v' & 0 & 0\\
 0 & 0 & 0 & 0
\end{array}
\right) \ .\label{sigma} 
\end{equation}
The constants of proportionality (\ref{rapporti}) in this particular fluid can then be inserted into the general formula (\ref{Tfluido}) for $T^{\mu\nu}$ to express the stress tensor in terms of the temperature $T$ and the velocity $v$
\beq
T^{\mu\nu}=\frac{\pi^3T^4}{16G_N}\left(
\begin{array}{cccc}
 3 & 0 & 4v & 0 \\
 0 & 1  & -\frac{v\p}{\pi T} & 0 \\
 4v & -\frac{v\p}{\pi T} & 1 & 0\\
 0 & 0 & 0 & 1
\end{array}
\right) \ .
\eeq

The velocity only depends on the coordinate $x$.  Let us choose boundary conditions so that the temperature $T$ also only depends on $x$.  Then the equations of motion (\ref{edm}) are simply
\beq
0=\partial_x T^{x\nu}.
\eeq
However all of the components $T^{x\nu}$ are constants except for $T^{xx}$ and $T^{xy}$.  Therefore the only nontrivial equation of motion at linear order in $v$ is
\beq
0=\partial_x T^{xx}+\partial_x T^{xy}=\frac{\pi^3T^3}{4G_N}T\p-\frac{\pi^2}{16G_N}\partial_x (T^3 v\p).
\eeq
When $v$ is linear in $x$, its second derivative vanishes.  Therefore this equation of motion, the conservation of momentum in the $y$ direction, implies that the temperature is constant.  More precisely it implies that $(\partial T)/T$ is negligible in this approximation.  In light of the relations (\ref{rapporti}) this is consistent with the incompressibility assumption that we have imposed on our fluid.  

Therefore we recover the fact that at sufficiently small velocities, a constant temperature and a $y$-velocity which is linear in $x$ solve the equations of motion (\ref{edm}).  Intuitively this is clear.  Further to the right, the fluid is moving faster.  Therefore the viscous force on a unit of fluid exerted by the faster fluid on its right (in the $+x$ direction) will accelerate it in the $+y$ direction, whereas the slower fluid on its left will exert a viscous force that decelerates it.  A steady flow occurs when these two forces cancel, which implies that the second derivative of the flow vanishes.  Had there been a temperature gradient, then the viscosity to density ratio would also have been stronger on one side by (\ref{rapporti}), and so this balance could only be maintained by introducing a second derivative of the velocity. 

Clearly a linear velocity also satisfies the nonrelativistic Navier-Stokes equation for an incompressible, Newtonian fluid
\beq
\rho(\partial_t v_k + (v\cdot\partial) v_k)=-\partial_k p + \nu \nabla^2 v_k. \label{nonrel}
\eeq 
In fact, each term vanishes independently.  Note that the vanishing of the $\partial p$ term is not merely a consequence of incompressibility.  In incompressible flows it may be of the same order as the viscous term.  It vanishes in this case because this provides a solution to (\ref{nonrel}) and it is consistent with the various nonrelativistic, small gradient and incompressible limits taken above.

In conclusion, we have considered fluid flows with an $x$-dependent velocity $v$ in the $y$ direction.  We have verified that, to linear order in $v$, these satisfy both the relativistic and nonrelativistic equations of motion when $v$ is linear in $x$ and the temperature $T$ is constant.  Our flows of interest will have $v=0$ on the left, $v$ linear on the right and an interpolating region inbetween.  Thus the equations of motion will be satisfied on the left and the right but not in the interpolating region, leading to a dual gravity solution which solves Einstein's vacuum equations on the left and right, but inbetween requires material described by a nontrivial stress tensor.  We have also found formula (\ref{proj}) and (\ref{sigma}) for the projector $P^{\mu\nu}$ and the shear tensor $\sigma^{\mu\nu}$ for general functions $v$, and so these results may be applied to the interpolating region.

Clearly if the fluid velocity is linear over a large enough distance, it will eventually approach the speed of light and the nonrelativistic approximation will break down.  Therefore our analysis is only relevant near the boundary.  The solution may be made global by introducing a second boundary, such that the velocity is constant on the other side of the second boundary.  We will see below that the stress tensor on the second boundary, to linear order in $v$, will be minus the stress tensor of the first boundary.  Of course at higher orders one may expect an attraction between the two boundaries, and so this solution will not be stationary.  The underlying assumption in this note is that the walls on the gravity side are built of a solid which remains fixed.  The fact that the null energy condition is violated may be a sign that such a material would be inconsistent, as some null energy violating configurations lead to superluminal propagation or instabilities \cite{superlum}, although some do not \cite{tranquilli}, in which case this configuration should only be considered over a sufficiently short timescale.  It may be that this timescale is never sufficient for turbulence to develop.





\section{The Gravity Dual} \label{gravsez}

The AdS/hydrodynamics correspondence yields a black brane metric dual to arbitrary flows in very particular conformal fluids, which for example obey the relations (\ref{rapporti}).  If the flow satisfies the hydrodynamic equations of motion (\ref{edm}), the dual satisfies the vacuum Einstein equations, in this case with a negative cosmological constant.  

\subsection{A note on ultralocality}

The correspondence, at least in the incarnation in Refs.~\cite{BHMR} and \cite{arbdim}, is ultralocal.  This means the following.  Consider the set of ingoing null geodesics which run from the boundary to the black brane horizon.  Clearly each point in the bulk is on precisely one such geodesic.  Also each point on the boundary is on one such geodesic.  Therefore these geodesics can be used to associate a fixed single boundary point to each bulk point.  This association is not one to one, there is an entire geodesic worth of bulk points associated to each boundary point.  

To implement the map, one identifies the boundary with the Minkowski space on which the fluid lives.  The metric and its derivatives at a point in the bulk are determined entirely by the fluid velocity and temperature and their derivatives at the associated boundary point.  One does not need to know the behavior of the fluid elsewhere.  This is the ultralocality of the correspondence.  In particular, the fact that the fluid satisfies the hydrodynamic equations of motion on the left and right (at $x<0$ and $x>d$) implies that the dual metric will satisfy the vacuum Einstein equations on the left and right, so long as the characteristic distance over which these quantities vary is greater than the inverse temperature.

Our fluid does not satisfy the hydrodynamic equations of motion at $0<x<d$.  This means that the dual gravitational configuration will not satisfy the vacuum Einstein equations, instead it will only satisfy Einstein's equations with a nonzero stress tensor, which we will calculate in Sec.~\ref{tsez}.

This ultralocality is somewhat different from the ultralocality that one encounters in classical field theories, or in the BKL limit of gravity theories, in that the bulk geometry is ultralocal in terms of null and not temporal evolution.  This ingoing null identification identifies the temporal evolution of the fluid with outward radial evolution for the gravity theory.  That is to say, the metric at larger radii but the same time is determined by the fluid in the future but at the same location.  In particular, a timeslice of the bulk geometry is determined by the evolution of the boundary fluid during a fixed interval of time.  This interval is of order the inverse temperature, and so no appreciable evolution may occur during this interval if the temperature is large enough for the correspondence to hold.  In this sense any fixed timeslice of the gravity dual contains only as much information as a fixed timeslice of the fluid, despite being one dimension greater.


As each event in the fluid is identified with an inward null geodesic in the bulk, the metric corresponding to this event appears to be falling towards the black hole at the speed of light.  This is not at all to say that there is a Killing vector in the inward null direction, the metric changes in that direction, but in a fashion which is fixed by the map.  Thus a disturbance on the boundary creates gravity waves which fly inward at the speed of light to the horizon.  Similarly, pasting together an infinite sequence of bulk timeslices which are separated by time intervals $1/T$, one obtains a pattern which falls from the boundary in to the horizon at the speed of light.  Although each individual timeslice is too small to see any evolution, the entire pattern is dual to the entire history of the flow.  Like a movie reel, the pattern in turn allows one to reconstruct the gravity dual, as it contains the timeslices.

One may use this identification to speculate on the gravitational dual of decaying turbulence.  For example, the inverse cascade of decaying (2+1)-dimensional turbulence consists of a chaotic period during which the fluid is subjected to random external forces followed by a relaxation period, characterized by the merging of well-separated vortices \cite{Kraichnan, mcwilliams}.  This would then be dual to a kind of forest of gravity waves falling from the boundary to the horizon, beginning when the boundary is subjected to a random perturbation.  First the canopy, representing the chaotic period, falls out of the boundary into the horizon.  When the external perturbation is turned off, it is followed by the branches representing the vortices, and then the branches merge into trunks as the vortices merge.  The usual cascade \cite{Kol1,Kol2} in (3+1)-dimensional turbulence may be similarly described, but when the boundary is randomly perturbed, the trunk falls out first.  This leads to the rather bizarre observation that black brane geometries in AdS$_4$ and AdS$_5$ respond very differently to random perturbations of their boundaries.  Needless to say, it would be interesting to make this picture precise, or to see whether it is inconsistent with the various approximations involved in the duality.

\subsection{The metric}

We will now calculate the metric dual to the flow (\ref{u}) using the map in Ref.~\cite{BHMR} with the simplified notation of Refs.~\cite{arbdim,rang}.  This map takes regions in which the flow satisfies the fluid equations to regions in which the metric satisfies the source-free Einstein equations.  Acting on the region in which the fluid does not satisfy the hydrodynamic equation, the map is not known to have any special properties other than continuity, which will produce an interpolation between the vacuum Einstein metrics on the two sides.  Therefore the choice of this map corresponds to a rather arbitrary choice of interpolation.  However we will see that this interpolation has two nice properties.  First, it is reasonably independent of the interpolating velocity function chosen.  In particular, the third and higher derivatives of the velocity will yield contributions to the integrated stress tensor which are suppressed by powers of $dT$, while the leading contribution is independent of $d$.  Second, the resulting stress tensor is simpler than the Israel stress tensor, it will have zero stress, whereas Israel's stress tensor has shear stress.

Ultralocality in the ingoing null direction implies that the simplest coordinates in which to express the metric are Gaussian null coordinates, in which $r$ parametrizes the ingoing null lines.  In these coordinates, the bulk 5-dimensional metric corresponding to an $x$-dependent 4-dimensional fluid flow is \cite{arbdim}
\begin{equation}
ds^2 =  G_{MN} dX^M dX^N = - 2 u_\mu(x) \, dx^\mu \, (dr + \mathcal{V}_\nu(r, x) \, dx^\nu) + \mathfrak{G}_{\mu \nu}(r, x) dx^\mu dx^\nu \label{met}
\end{equation}
where, up to second derivatives in $v$, $ \mathcal{V}_\nu$ and $ \mathfrak{G}_{\mu \nu}$ are defined as
\begin{multline}
\mathcal{V}_\nu = r \mathcal{A}_\mu - \mathcal{S}_{\mu \lambda} u^\lambda - {\mathfrak v}_1(br) P_{\mu}^\nu \mathcal{D}_\lambda \sigma^\lambda_\nu \\
+ u_\mu \left[\frac{1}{2} r^2 \left(1-\frac{1}{(br)^4}\right) - \frac{1}{4(br)^4}\omega_{\alpha \beta} \omega^{\alpha \beta} + {\mathfrak v}_2(br) \frac{\sigma^{\alpha \beta} \sigma_{\alpha \beta}}{d-1} \right] \label{vdef}
\end{multline}
and
\begin{multline}
\mathfrak{G}_{\mu \nu} = r^2 P_{\mu \nu} - \omega_\mu^\lambda \omega_{\lambda \nu} + 2 (br)^2 {\mathfrak g}_1(br)\left[\frac{1}{b}\sigma_{\mu \nu} + {\mathfrak g}_1(br) \sigma_{\mu}^\lambda \sigma_{\lambda \nu} \right] - {\mathfrak g}_2(br)  \frac{\sigma^{\alpha \beta} \sigma_{\alpha \beta}}{d-1} P_{\mu \nu} \\
- {\mathfrak g}_3 (br) \left[{\mathfrak T}_{1 \mu \nu} + \frac{1}{2} {\mathfrak T}_{3 \mu \nu}+ 2 {\mathfrak T}_{2 \mu \nu} \right] + {\mathfrak g}_4(br)[{\mathfrak T}_{1 \mu \nu} +{\mathfrak T}_{4 \mu \nu} ] \ . \label{gdef}
\end{multline}
The functions that appear in this definition are defined in Ref.~\cite{arbdim}, we will quote the definitions as they are needed.

Eq.~(\ref{met}) is the bulk metric dual to a fluid flow in an arbitrary curved space.  The fluid is conformally invariant, and the conformal invariance has been used to write the metric in a compact form using objects which transform covariantly under the conformal symmetry.  We are interested in a flat boundary, and so many of these objects will vanish.  In fact, since $u$ only depends on $x$, but only has nonvanishing components in the $t$ and $y$ directions, even the gauge field for a Weyl transformation will vanish
\begin{equation}
\mathcal{A}_\mu = u^\lambda \nabla_\lambda u_\mu - \frac{1}{d-1}u_\mu \nabla^\lambda u_\lambda=0 
\end{equation}
as $u^\lambda \nabla_\lambda u_\mu=0$ and $ \nabla_\lambda u^\lambda =0$. This implies that the Weyl-covariant derivative reduces to the ordinary derivative
\begin{equation}
\mathcal{D}_\mu  = \partial_\mu \ .
\end{equation}

The Weyl-covariant Schouten tensor $\mathcal{S}$ is proportional to the Weyl-covariant curvature of the boundary.  As the Weyl-covariant derivative is just the ordinary derivative, this is just the ordinary curvature.  As the boundary is Minkowski space, the curvature vanishes, and so the Weyl-covariant Schouten tensor also vanishes
\begin{equation}
\mathcal{S}_{\mu \nu} = 0 \ .
\end{equation}
Similarly the Weyl-covariant Weyl curvature $\mathcal{C}$ is the sum of the ordinary Weyl curvature and the curvature of the Weyl tensor, which both vanish and so
\begin{equation}
\mathcal{C}_{\mu \nu \lambda \sigma} = 0\ .
\end{equation}

The vorticity $\omega$ does not vanish, however like the shear strain $\sigma$ it is of first order in $v$.  Therefore $\omega^2$, $\omega\sigma$ and $\sigma^2$ terms are all of order $\mathcal{O}(v^2)$ and so do not contribute at order $\mathcal{O}(v)$.  Thus only the third and fourth terms of (\ref{vdef}) contribute to $\mathcal{V}_\mu$.  The third term is easily evaluated
\begin{equation}
\mathcal{D}_\lambda \sigma^\lambda_\nu=\partial_\lambda \sigma^\lambda_\nu =  \partial_x \sigma^x_\nu
\simeq \left(
\begin{array}{c}
 0\\
 0\\
 \frac{1}{2} v''\\
 0
\end{array}
\right) \ .
\end{equation}
Adding the third and fourth terms one finds $\mathcal{V}_\mu$ at order $\mathcal{O}(v)$
\begin{equation}
\mathcal{V}_\mu \simeq {\mathfrak v}_1(br) 
 \left(
\begin{array}{c}
 0\\
 0\\
 \frac{1}{2} v''\\
 0
\end{array}
\right) + 
 \left(
\begin{array}{c}
 1\\
 0\\
 v\\
 0
\end{array}
\right) \frac{1}{2}\left(r^2-\frac{1}{b^4r^2}\right). \label{v2}
\end{equation}

The functions ${\mathfrak T}$ are easily expressed in terms of the shorthand notation $<>$, which symmetrizes and contracts with the projectors $P_{\mu \nu}$.  The first two are identically zero, while the second two are of order $\mathcal{O}(v^2)$
\begin{align}
{\mathfrak T}_{1 \mu \nu} &= 2 u^\alpha \mathcal{D}_\alpha \sigma_{\mu \nu} = 0 \\
{\mathfrak T}_{2 \mu \nu} &= C_{\mu \alpha \nu \beta} u^\alpha u^\beta = 0\\
{\mathfrak T}_{3 \mu \nu} &= 4 \sigma^{\alpha \langle \mu} \sigma^{\nu\rangle}_\alpha \sim 0 \\
{\mathfrak T}_{4 \mu \nu} &= 4 \sigma^{\alpha \langle \mu} \omega^{\nu\rangle}_\alpha \sim 0. 
\end{align}
Therefore only the first and third terms of (\ref{gdef}) contribute to $\mathfrak{G}_{\mu \nu}$
\begin{equation}
\mathfrak{G}_{\mu \nu} \simeq  
r^2 \left(
\begin{array}{cccc}
 0 & 0 & v & 0 \\
 0 & 1  & 0 & 0 \\
 v & 0 & 1 & 0\\
 0 & 0 & 0 & 1
\end{array}
\right) 
+  b r^2 {\mathfrak g}_1(br)  
\left(
\begin{array}{cccc}
 0 & 0 & 0 & 0 \\
 0 & 0  &  v' & 0 \\
 0 & v' & 0 & 0\\
 0 & 0 & 0 & 0
\end{array}
\right) \ . \label{g2}
\end{equation}

Finally, inserting Eqs.~(\ref{v2}) and (\ref{g2}) into (\ref{met}), one finds the final form of the metric
\begin{multline}
ds^2 = - r^2 \left(1-\frac{1}{b^4 r^4}\right)dt^2 - 2 dt dr  + \left(2 \frac{v}{b^4 r^2}+{\mathfrak v}_1(br) v'' \right) dt dy \\
 +r^2(dx^2 + dy^2+dz^2)+ 2 r^2 b\, {\mathfrak g} _1(br) v' dx dy - 2 v dy dr
\end{multline}
where $b=1/{\pi T}$ and the functions ${\mathfrak v}_1$ and ${\mathfrak g}_1$ are defined in Eqs.~(\ref{v1}) and (\ref{g1}).  Using the basis $(t,x,y,z,r)$, we may write the metric in matrix form as
\[
g_{\mu \nu}=
\begin{pmatrix}
-\left(r^2-\frac{1}{b^4 r^2}\right) & 0 & \frac{v}{b^4 r^2}+\frac{1}{2}{\mathfrak v}_1(br) v''  & 0 & -1\\
0 & r^2 & r^2 b\, {\mathfrak g} _1(br) v'  & 0 & 0\\
 \frac{v}{b^4 r^2}+\frac{1}{2}{\mathfrak v}_1(br) v''  &  r^2 b\, {\mathfrak g} _1(br) v'  & r^2 & 0 & - v\\ 
 0 & 0 & 0 & r^2 & 0\\
 -1 & 0 & -v & 0 & 0
\end{pmatrix} \   \label{metfin}
\]
while the inverse metric is
\[
g^{\mu \nu}=
\begin{pmatrix}
 0 & 0 & -\frac{v}{r^2} & 0 & -1\\
 0 & \frac{1}{r^2} & - \frac{b\, {\mathfrak g} _1(br) v'}{r^2} & 0 & 0\\
 -\frac{v}{r^2} & - \frac{b\, {\mathfrak g} _1(br) v'}{r^2} & \frac{1}{r^2} & 0 & \frac{{\mathfrak v}_1(br) v''}{2 r^2} + v\\
 0 & 0 & 0 & \frac{1}{r^2} & 0\\
 -1 & 0 & \frac{{\mathfrak v}_1(br) v''}{2 r^2} + v & 0 & r^2 - \frac{1}{b^4 r^2}
\end{pmatrix} \ .
\]
 
\subsection{Christoffel symbols}
  
In Sec.~\ref{tsez} we will see that the leading contribution to the stress tensor comes from the second derivative of the velocity.  Contributions at that order come from the second derivative of the velocity in the curvature, which in turn contains contributions from first and second derivatives of the velocity in the Christoffel symbols, as well as from Christoffel symbols which are velocity independent as these are multiplied by velocity-dependent terms when calculating the curvature.  We will now calculate all of the Christoffel symbols up to first order in $v$, $v\p$ and $v\p\p$, although the $v$ terms will not contribute to the stress tensor.

We will begin with the terms at order $\mathcal{O}(v^0)$, these are just the Christoffel symbols of the static black brane 
\begin{align}
\Gamma_{tt}^t &= -\left(r+\frac{1}{b^4r^3}\right), & \Gamma_{tt}^r &=r^3 - \frac{1}{b^8 r^5}, & \nonumber\\
\Gamma_{xx}^t &= \Gamma_{yy}^t = \Gamma_{zz}^t = r, & \Gamma_{tr}^r &=  \Gamma_{rt}^r= r + \frac{1}{b^4 r^3}, &\nonumber\\
\Gamma_{xr}^x &= \Gamma_{rx}^x = \frac{1}{r}, & \Gamma_{yr}^y &=\Gamma_{ry}^y= \frac{1}{r}, & \nonumber\\
\Gamma_{zr}^z &= \Gamma_{rz}^z = \frac{1}{r}, & \Gamma_{xx}^r &= \Gamma_{yy}^r = \Gamma_{zz}^r = -\left(r^3-\frac{1}{b^4 r}\right). \label{Chris0}
\end{align}

The new terms, at order $\mathcal{O}(v)$, are
\begin{align}
\Gamma_{ty}^t &= \Gamma_{yt}^t = \frac{1}{4}b\,{\mathfrak v}'_1(br) v'' - \frac{v}{b^4 r^3}, \hspace{4.2cm} 
\Gamma_{yr}^x = \Gamma_{ry}^x = \frac{v'}{2 r^2}+\frac{b^2}{2} {\mathfrak g}_1(br) v'\nonumber\\
\Gamma_{tt}^y &= \left(r+ \frac{1}{b^4 r^3}\right)\left(\frac{{\mathfrak v}_1(br) v''}{2 r^2} + v\right) ,\hspace{3.5cm}\Gamma_{tx}^y = \Gamma_{xt}^y = \frac{v'}{2 b^4 r^4}\nonumber\\
\Gamma_{xx}^y &= b {\mathfrak g}_1(br) v'' - vr- \frac{{\mathfrak v}_1(br) v''}{2 r}
,\hspace{4cm}
\Gamma_{tr}^y = \Gamma_{rt}^y = \frac{v}{r} + \frac{1}{4}\frac{{\mathfrak v}'_1(br) b v''}{r^2} \nonumber\\
\Gamma_{xr}^y &= \Gamma_{rx}^y = \frac{1}{2} b^2 {\mathfrak g}'_1(br) v' - \frac{v'}{2 r^2},\hspace{4.4cm}
\Gamma_{yy}^y = \Gamma_{zz}^y = - r v - \frac{{\mathfrak v}_1(br) v''}{2r}\nonumber\\
\Gamma_{ty}^r &= \Gamma_{yt}^r = \left(r^2-\frac{1}{b^4 r^2}\right)\left(\frac{v}{b^4 r^3}-\frac{1}{4} b {\mathfrak v}'_1(br) v'' \right),\hspace{1.3cm}
\Gamma_{ty}^x = \Gamma_{yt}^x  =-\frac{v'}{2 b^4 r^4} \nonumber\\
\Gamma_{yr}^r &= \Gamma_{ry}^r = \frac{v}{b^4 r^3} - \frac{1}{4} b {\mathfrak v}'_1(br) v'' + \frac{{\mathfrak v}_1(br) v''}{2 r}+ v r,\hspace{1.4cm} 
\Gamma_{yr}^t = \Gamma_{ry}^t  =-\frac{v}{r}\nonumber
\end{align}

\begin{align}
\Gamma_{xy}^r &= \Gamma_{yx}^r = -\frac{1}{2} r^2 v' - \left(r^2-\frac{1}{b^4 r^2}\right)\left(r b \, {\mathfrak g}_1(br) v' + \frac{1}{2} r^2 b^2 \, {\mathfrak g}'_1(br) v' \right)  \nonumber\\
\Gamma_{xy}^t &= \Gamma_{yx}^t = \frac{1}{2}\left(v'+2 r b\,{\mathfrak g}_1(br)v'+r^2 b^2 \,{\mathfrak g}'_1(br) v' \right)  . \label{Chris1}
\end{align}

\subsection{The Riemann tensor and the Ricci tensor and scalar}

Using the Christoffel symbols one can now easily compute the Riemann tensor.  The order $\mathcal{O}(v^0)$ terms again are just those of the static AdS black brane
\begin{align}
R_{trtr} &=  1 - \frac{3}{b^4 r^4} \\
R_{txtx} &= R_{tyty} = R_{tztz} = r^4 - \frac{1}{b^8 r^4} \\
R_{txrx} &= R_{tyry} = R_{tzrz} = r^2 + \frac{1}{b^4 r^2} \\
R_{xyxy} &= R_{xzxz} = R_{yzyz} = - r^4 + \frac{1}{b^4}. 
\end{align} 

The bulk stress tensor is entirely determined by the contributions to the Riemann tensor which do not solve the fluid equations of motion, as it is these that do not solve the vacuum Einstein equations.  If $v$ is a constant, this yields the boosted black brane, which satisfies the vacuum Einstein equations.  If $v$ is linear, then again this is a solution of the linear order fluid equations as we have checked above, and therefore as we will check below $v\p$ will not contribute to the gravitational stress tensor at order $\mathcal{O}(v)$.  Therefore the first nontrivial contributions to the stress tensor arise from the Riemann tensor at linear order in $v''$ ($v= v' = 0$) 
\begin{align}
R_{trty} &= - \frac{1}{2}\left(1+ \frac{1}{b^4 r ^4}\right) {\mathfrak v}_1(br) v'' \\
R_{xrxy} &= \frac{v''}{4}\left(2+ 2{\mathfrak v}_1(br)+2b^2 r^2\, {\mathfrak g}'_1(br)- b r {\mathfrak v}'_1(br)\right)\  \\
R_{zrzy} &=  \frac{v''}{4}\left(2{\mathfrak v}_1(br)- b r {\mathfrak v}'_1(br)\right)\  \\
R_{yxtx} &= - \frac{v''}{4 b^4 r^2}(2+(b^5 r^5- br){\mathfrak v}'_1(br)) \	  \\
R_{yrtr} &=  \frac{v''}{4}\left({\mathfrak v}'_1\,\frac{b}{r} - b^2 {\mathfrak v}''_1 \right) .  
\end{align}

As a check on our calculation, we will also calculate the contributions to the various tensors at linear order in the nondifferentiated velocity $v$
\begin{align}
R_{txyx} &= R_{tzyz} = \frac{1}{b^4}\left(1- \frac{1}{b^4 r^4}\right) v \  \\
R_{tytr} &= - \left(r^2+ \frac{1}{b^4 r^2}\right) v \  \\
R_{tryr} &= \left(1- \frac{3}{b^4 r^4}\right) v \   \\
R_{xyxr} &= R_{zyzr} = \left(r^2 + \frac{1}{b^4 r^2}\right) v 
\end{align}
and in $v\p$
\begin{align}
R_{txty} &= (b^3 r^3 + b^7 r^7 + (b^8 r^8 -1)(2 {\mathfrak g}_1(br)+ br {\mathfrak g}'_1(br))) \frac{v'}{2 b^7 r^4} \  \\
R_{txyr} &=  (-2 + b^4 r^4 + (b^5 r^5 + b r)(2 {\mathfrak g}_1(br)+ br {\mathfrak g}'_1(br))) \frac{v'}{2 b^4 r^3} \  \\
R_{tyxr} &=  - (2 + b^4 r^4 + (b^5 r^5 + b r)(2 {\mathfrak g}_1(br)+ br {\mathfrak g}'_1(br))) \frac{v'}{2 b^4 r^3} \  \\
R_{trxy} &= - \frac{2 v'}{b^4 r^3} \  \\
R_{xzyz} &= - (b^3 r^3 + (b^4 r^4 -1)(2 {\mathfrak g}_1(br)+ br {\mathfrak g}'_1(br))) \frac{v'}{2 b^3} \  \\
R_{xryr} &= - \frac{1}{2} b^2 r v' (2 {\mathfrak g}'_1(br)+ br {\mathfrak g}''_1(br)) \ .
\end{align}


The Ricci tensor is now easily calculated.  Again the order $\mathcal{O}(v^0)$ terms are those of the static black brane solution
\begin{align}
R_{tt} &= 4 r^2 - \frac{4}{b^4 r^2}  \\
R_{tr} &= R_{rt} = 4 \\
R_{xx} &= R_{yy} = R_{zz} = - 4r^2 \ .
\end{align}

Contributions to the stress tensor will arise from the $v''$ terms, ($v= v' = 0$) 
\begin{align}
R_{ty} &= R_{yt} = -\frac{v''}{4 b^4 r^4}(2 + 4(1+ b^4 r^4) {\mathfrak v}_1(br)+ (b^5r^5 -br)({\mathfrak v}'_1(br)+b r{\mathfrak v}''_1(br))) \\
R_{ry} &= R_{yr} = \frac{v''}{4 r^2}(2+ 4 {\mathfrak v}_1  + br(2 br \,{\mathfrak g}'_1(br)- {\mathfrak v}'_1(br)+ b r {\mathfrak v}''_1(br) ) \ .
\end{align}

The terms in the Ricci tensor proportional to $v$ are those of a rigidly boosted black brane
\beq
R^{(v)}_{ty} = - \frac{4 v}{b^4 r^2} \hsp
R^{(v)}_{yr} = 4 v \ 
\eeq
which provides an exact solution both to the hydrodynamic equations and to Einstein's equations with a negative cosmological constant.  Again the terms linear in $v'$ yield a solution to the fluid equations and so Einstein's equations, although only to linear order $\mathcal{O}(v)$
\beq
R_{xy} = - (8 b^3 r^3 {\mathfrak g}_1(br)+ (5 b^4 r^4 -1){\mathfrak g}'_1(br))\frac{v'}{2 b^2 r} \ .
\eeq

Using the large $r$ asymptotic expansions of Ref.~\cite{arbdim} 
\begin{align}
{\mathfrak g}_1 &\sim \frac{1}{b r} - \frac{1}{4 b^4 r^4} + \dots \  \\
{\mathfrak v}_1 &\sim - \frac{1}{12 b^4 r^4}+ \frac{2}{5 b^3 r^3} + \dots \ 
\end{align}
we find that the asymptotic behaviors of the $v\p\p$ terms in the Ricci tensor are
\begin{align}
R_{ty} &\sim \frac{13}{10} \frac{v''}{b^3 r^3} \  \\
R_{ry} &\sim \frac{1}{5} \frac{v''}{b^2 r^4} \ . 
\end{align}


The Ricci scalar is
\begin{equation}
R = -20 \ .
\end{equation}
There is no contribution at order $\mathcal{O}(v)$ to the Ricci scalar.  This is guaranteed for any solution of the vacuum Einstein equations with cosmological constant $\Lambda=-6$, and so there could not have been any corrections from the $v$ and $v\p$ terms.  There are no corrections from the $v\p\p$ terms at linear order because the corresponding components of the inverse metric are themselves of order $\mathcal{O}(v)$, and so the contributions to the Ricci tensor are of order $\mathcal{O}(v^2)$.

\subsection{Contributions to the Ricci tensor at $\mathcal{O}(v^{(3)})$ and $\mathcal{O}(v^{(4)})$}

Before continuing with the calculation of the bulk stress tensor, we will pause to discuss some of the approximations that we have made.  We have made two truncations.  First, we have calculated everything at order $\mathcal{O}(v)$.  As we are working in units in which $c=1$, $v$ is small for nonrelativistic speeds and so this is a valid approximation in a region in which the flow is sufficiently slow.

A more dangerous truncation is that of higher derivatives of the velocity.  The gravity/hydrodynamics correspondence is a one to one map between gravitational and fluid solutions in a derivative expansion.  More precisely, the $k$th order map relates the truncation of the fluid equations to $k$ derivatives and that of the gravity equations to $(k+1)$ derivatives.  The iterative procedure described in Ref.~\cite{BHMR} in principle determines this map for all $k$, however in practice this map has only been determined to order $k=2$.  In other words, it provides a metric as a function of $v$, $v\p$ and $v\p\p$, however a perfect matching with Einstein's equations would require also corrections involving the higher derivatives $v^{(k)}$ which are not known.

General arguments based on dimensional analysis suggest that these corrections become smaller at higher $k$.  In general one expects that each derivative leads to a contribution which is subdominant by a factor of $Tl$ with respect the previous derivative, where $l$ is the distance scale of the derivative.  Ideally one would like to check this claim for all terms with, say, three or four derivatives.  However this would require a knowledge of the map at orders $k=3$ and $k=4$.

The map at order $k=2$, which we have used, does produce some terms in the curvature which depend on the third and fourth derivatives of $v$.  In this subsection we will verify that two of these have the expected convergence scaling, and determine the corresponding condition on our fluid flow.  In other words, we determine a necessary condition for the derivative expansion to apply to our flow.

The Ricci tensor components $R_{xy}$ and $R_{ty}$ have corrections from the third and fourth derivatives of the velocity respectively 
\begin{align}
R_{xy}^{(3)} &= -\frac{v^{(3)}(x) \left(b r {\mathfrak v}'_1(b r)+{\mathfrak v}_1(b r)\right)}{4 r} \  \\
R_{ty}^{(4)} &= -\frac{v^{(4)}(x) {\mathfrak v}_1(b r)}{4 r^2} \ .
\end{align}
We want to determine the condition under which $R_{ty}^{(4)}$ is subdominant to $R_{xy}^{(3)}$.  As the higher derivatives of $v$ define an interpolating function between two solutions over an interval of length $d$, each derivative is larger than the previous one by about $1/d$.  In other words, $\partial_x\sim 1/d$.  

To test the subdominance of $R_{ty}^{(4)}$, it is sufficient to compare it to the similar term in $R_{xy}^{(3)}$, which contains ${\mathfrak v}_1$.  The ratio of these terms is
%
\begin{equation}
\frac{R_{ty}^{(4)}}{R_{xy}^{(3)}} \sim \frac{v^{(4)}(x)}{rv^{(3)}(x)} \sim \frac{1}{r d}
\end{equation}
therefore the fourth order term is subdominant if $d\gg1/r$ in the entire bulk.  The bulk extends from the horizon at $r=1/b=\pi T$ to the boundary at $r=\infty$.  Therefore convergence requires
\beq
d\gg\frac{1}{\pi T}. \label{spesso}
\eeq
This fourth order term is suppressed by $\pi d T$ with respect to the third order term, in line with the above expectations from dimensional analysis.  This means that the gravity duality procedure is only convergent when $d$ is sufficiently large.  Of course, the duality never yields a solution of the vacuum Einstein equations, and so one may argue that its convergence is immaterial.  Nonetheless, it is only well-defined as a series when $d$ satisfies (\ref{spesso}).

\subsection{The static black brane solution}

As a check on our calculation and conventions, we recover that the static $(v=0)$ black brane satisfies the vacuum Einstein equations with cosmological constant $\Lambda=-6$
\begin{equation}
R_{\mu \nu} - \frac{1}{2} R g_{\mu \nu} = \left(
\begin{array}{ccccc}
 \frac{6}{b^4 r^2}-6 r^2 & 0 & 0 & 0 & -6 \\
 0 & 6 r^2 & 0 & 0 & 0 \\
 0 & 0 & 6 r^2 & 0 & 0 \\
 0 & 0 & 0 & 6 r^2 & 0 \\
 -6 & 0 & 0 & 0 & 0
\end{array}
\right) \ . \label{bn}
\end{equation}

\section{Two Calculations of the Stress Tensor} \label{tsez}

In this section we will calculate the bulk stress tensor of the surface layer interpolating between the vacuum gravity solutions using two different methods, corresponding to two different metrics.  First, we will apply the duality map of Ref.~\cite{BHMR} to a fluid flow which interpolates between the two solutions, the stationary solution on the left and the linear velocity solution on the right.  In this case, as we have seen, the interpolating region is necessarily larger than the inverse temperature.  Next, we will directly interpolate between the gravitational solutions using the Israel matching conditions \cite{Israel}.  This method requires the interpolating region to be very thin, and uses the assumption that in this limit the extrinsic curvature remains bounded.

\subsection{Interpolating between the hydrodynamic flows} \label{T1sez}

The duality map of Ref.~\cite{BHMR} takes a fluid flow and yields a dual metric.  This dual metric solves the vacuum Einstein equations when the fluid flow satisfies the hydrodynamic equations of motion (\ref{edm}).  If the flow does not satisfy the equations of motion, the dual metric does not satisfy the vacuum Einstein equations.  Thus apparently there is no benefit in using this map over any other map.  However we will use the map, and observe the consequences.  The resulting dual metric will necessarily solve Einstein's equations with some value of the stress tensor
\begin{equation}
8\pi G_N T_{\mu \nu} = R_{\mu \nu} - \frac{1}{2} R g_{\mu \nu} + \Lambda g_{\mu \nu} \ . \label{ein}
\end{equation}
We will determine this value. 

We saw in Eq.~(\ref{bn}) that there is no contribution to the stress tensor at order $\mathcal{O}(v^0)$.  We have argued that, at order $\mathcal{O}(v)$, the dominant contributions to the stress tensor are proportional to $v\p\p$.  These are easily found from (\ref{ein}) to be 
\begin{align}
T_{ty} &= \frac{v''(x) \left(4 \left(b^4 r^4-1\right) {\mathfrak v}_1(b r)-b r \left(b^4 r^4-1\right) \left({\mathfrak v}'_1(b r)+b r {\mathfrak v}''_1(b
   r)\right)-2\right)}{32\pi G_N b^4 r^4} \, \\
T_{ry} &= \frac{v''(x) \left(4 {\mathfrak v}_1(b r)+b r \left(2 b r {\mathfrak g}'_1(b r)-{\mathfrak v}'_1(b r)-b r
   {\mathfrak v}''_1(b r)\right)+2\right)}{32\pi G_N r^2}\ .
\end{align}
There appears to also be a contribution proportional to $v\p$
\begin{equation}
T_{xy} = -\frac{v'(x) \left(b r \left(\left(b^4 r^4-1\right) {\mathfrak g}''_1(b r)+3 b r\right)+\left(5 b^4 r^4-1\right) {\mathfrak g}'_1(b r)\right)}{16\pi G_N b^2 r}. \label{txy}
\end{equation}
At order $v\p$ one expects no contributions to the stress tensor, as a solution with a linear velocity satisfies the fluid equations at order $\mathcal{O}(v)$.  Therefore a nontrivial contribution would be in contradiction with the gravity/hydrodynamics correspondence.  We will see shorty that this contribution is in fact equal to zero.

The functions ${\mathfrak v}_1(r)$ and ${\mathfrak g}_1(r)$ are defined as
\begin{align}
{\mathfrak v}_1(r) &= \frac{2}{r^2} \int_{r}^{\infty} dx \, x^3 \int_x^{\infty} dy \frac{y-1}{y^3(y^4-1)} \label{v1}\\
{\mathfrak g}_1(r) &= \int_{r}^{\infty} dx \, \frac{x^3-1}{x(x^4-1)} \ . \label{g1}
\end{align}
Integrating we \cite{arbdim} obtain analytical expressions for ${\mathfrak v}_1(r)$ and ${\mathfrak g}_1(r)$ 
\begin{align}
{\mathfrak v}_1 &= - \frac{1}{4} + \frac{r}{2} + \frac{1}{8 r^2}(r^4-1)\left(\log{\frac{(r^2+1)}{(r+1)^2}}+2 \tan ^{-1}(r)-\pi\right) \\
{\mathfrak g}_1 &= \frac{1}{4} \left(\log{\left(\frac{(1+r)^2(1+r^2)}{r^4}\right)}-2 \tan ^{-1}(r)+\pi \right) \  .
\end{align}
The derivatives of these expressions are
\begin{align}
{\mathfrak g}'_1 &= \frac{1}{2} \left(\frac{r}{r^2+1}-\frac{1}{r^2+1}+\frac{1}{r+1}-\frac{2}{r}\right)\  \label{gp} \\
{\mathfrak g}''_1 &= -\frac{r^2}{\left(r^2+1\right)^2}+\frac{r}{\left(r^2+1\right)^2}+\frac{1}{2(r^2+1)}+\frac{1}{r^2}-\frac{1}{2(r+1)^2} \  \label{gpp}
\end{align}
and
\begin{multline}
{\mathfrak v}'_1 = -\frac{1}{4 r^3} \left( \pi  r^4+2 \left(r^4+1\right) \log (r+1)-2 \left(r^4+1\right) \tan ^{-1}(r) \right. \\ 
\left. -4 r^3+2 r^2-\left(r^4+1\right) \log \left(r^2+1\right)+\pi\right) 
\end{multline}

\begin{multline}
{\mathfrak v}''_1 = -\frac{1}{4 r^4 (r+1) \left(r^2+1\right)}   \left(3 \log \left(r^2+1\right)+\pi  (r+1) \left(r^2+1\right) \left(r^4-3\right)  \right.\\
+\left(r (r+1) \left(r^4+r^2-3\right)-3\right) r \left(2 \log
   (r+1) \right. \\
     -\log\left.  \left(r^2+1\right)\right)-2 (r+1) \left(r^2+1\right) \left(r^4-3\right) \tan ^{-1}(r)\\
    \left. -2 \left(2 r^4+r^3+r^2+r+3\right) r^2-6 \log
   (r+1)\right) .
\end{multline}

The explicit formula Eqs.~(\ref{gp}) and (\ref{gpp}) for the derivatives of ${\mathfrak g}_1$ can be combined to show that
\begin{equation}
r \left(\left(r^4-1\right) {\mathfrak g}''_1(r)+3 r\right)+\left(5 r^4-1\right){\mathfrak g}'_1(r) = 0 \ .
\end{equation}
This combination is proportional to formula (\ref{txy}) for $T_{xy}$, therefore
\beq
T_{xy}=0
\eeq
and there are no contributions proportional to $v\p$.  

Similarly one may evaluate the combination of functions that appears in $T_{ry}$ 
\begin{align}
r \left(2 r {\mathfrak g}'_1(r)-r {\mathfrak v}''_1(r)-{\mathfrak v}'_1(r)\right)+4 {\mathfrak v}_1(r)+2 = 0 \ .
\end{align}
This implies that
\beq
T_{ry}=0
\eeq
leaving only $T_{ty}$, the momentum in the $y$ direction.  Thus the bulk stress tensor contains no stress, only momentum.

We may use the exact expressions for the functions ${\mathfrak g}_1$ and ${\mathfrak v}_1$ to simplify the only nonvanishing component of the stress tensor 
\begin{equation}
T_{ty}= -\frac{v\p\p (x)}{16\pi G_N b^3 r^3} \ . \label{tty}
\end{equation}
Using the fundamental theorem of calculus, this may be integrated over the interpolating region to obtain
\beq
\int_0^d dx\ T_{ty}= -\frac{v\p}{16\pi G_N b^3 r^3} 
\eeq
where $v\p$ is the derivative of the velocity in the region $x>d$.  In particular, at this leading order the integrated stress tensor of the surface layer is independent of the interpolation and independent of $d$.  Of course it still depends on the map that we used to generate the dual metric.

Had the $v\p$ term been the dominant contribution, the stress tensor would have been constant, and so the integral would be have proportional to $d$.  Similarly a $v^{(3)}$ term would have led to a stress tensor proportional to $1/d$, and higher powers of $v$ to other scalings.  Therefore it is somewhat nontrivial that the leading contribution to the integrated stress tensor is in fact $d$-independent.  Clearly this $d$-independence is desirable, as $d$ is not a physical quantity but merely an artifact of the scheme that we used to regularize the divergent second derivative of the fluid velocity.

The bulk stress tensor does not satisfy the null energy condition.  As the only nonvanishing component is $T_{ty}$, the only nonvanishing product of a null vector $w$ and the stress tensor is
\beq
w^\perp Tw=2w^t T_{ty} w^y. \label{prod}
\eeq
As $T_{ty}$ is already of order $\mathcal{O}(v)$, at order $\mathcal{O}(v)$ one need only consider the terms in $w$ of order $\mathcal{O}(v^0)$.  That is to say, $w$ only needs to be null with respect to the static black brane metric.  Consider for example the null vectors $w_\pm$
\beq
w^t_{\pm}=r\hsp
w^y_{\pm}=\pm r\sqrt{1-\frac{1}{b^4r^4}}. \label{w}
\eeq
The product (\ref{prod}) is 
\beq
w_\pm^\perp Tw_\pm=\mp\frac{v''(x)}{8\pi G_N b^3 r}\sqrt{1-\frac{1}{b^4r^4}} \label{prod2}
\eeq
which is nonzero.  However $w_+$ and $w_-$ yield opposite signs, as incidentally do the two choices of signs of $v$.  Therefore at least one of these will yield a negative product, and so the bulk stress tensor does not satisfy the null energy condition.  This may or may not mean that no external matter may be consistently added which produces such a surface layer.

 
 

\subsection{Israel's matching conditions on the gravity duals}

We will now calculate the bulk stress tensor in a different geometry.  Following Ref.~\cite{Israel}, we will consider the vacuum Einstein solution corresponding to a static fluid on the left and that corresponding to a linear velocity flow on the right.  These solutions will be glued together by interpolating continuously between the two metrics over a distance $d$ and taking the limit $d\rightarrow 0$ such that the extrinsic curvature remains bounded.  In Ref.~\cite{Israel}, Israel has shown that the resulting configuration contains two solutions separated by a surface layer whose bulk stress tensor is independent of the interpolation used.

Following Ref.~\cite{Israel}, the first step in the calculation of the stress tensor is the definition of the unit normal vector to the hyperplane
\begin{equation}
n_\mu = \{0, r, 0, 0, 0\} \ ,
\end{equation}
which satisfies the normalization condition
\begin{equation}
n_\mu g^{\mu \nu}n_\nu = \frac{1}{r^2}(n_x)^2 = 1 \ .
\end{equation}
The surface layer $\Sigma$ extends along all of the directions except for the $x$ direction.  A basis of tangent vectors to $\Sigma$ is
\begin{equation}
ds = e_{(i)} d x^i 
\end{equation}
where
\begin{align}
e_{(t)} &= \{1, 0, 0, 0, 0\} \  \\ 
e_{(y)} &= \{0, 0, 1, 0, 0\} \  \\
e_{(z)} &= \{0, 0, 0, 1, 0\} \   \\
e_{(r)} &= \{0, 0, 0, 0, 1\} \ .
\end{align}
In terms of these tangent vectors the extrinsic curvature may be calculated as
\begin{equation}
K_{ij} = e_{(j)}\cdot \nabla_j n = \frac{\partial n_j}{\partial x^i} - n^m \Gamma_{m,ji} = \frac{\partial n_j}{\partial x^i} - n_m \Gamma^m_{ji} \ . \label{K}
\end{equation}

On the left, where the fluid is static ($v = 0$), substituting (\ref{Chris0}) into (\ref{K}) one finds no extrinsic curvature $K^{(-)}$
\begin{align}
K_{ty}^{(-)} &= - r \Gamma^x_{ty} = 0 \  \\
K_{yr}^{(-)} &= - r \Gamma^x_{yr} =  0 \  .
\end{align}
On the right, where the fluid velocity is linear, the Christoffel symbols of Eq.~(\ref{Chris1}) yield a nontrivial extrinsic curvature $K^{(+)}$.
\begin{align}
K^{(+)}_{ty} &= - r \Gamma^x_{ty} = \frac{v'}{2 b^4 r^3} \  \\
K^{(+)}_{yr} &= - r \Gamma^x_{yr} =  \frac{v'}{2 r}\left(1 + b^2 r^2 {\mathfrak g}'_1(br) \right) \ . 
\end{align}

The tensor $\gamma_{ij}$ is defined to be the difference between the extrinsic curvatures on the two sides of the surface layer
\begin{equation}
\gamma_{ij} = K^{(+)}_{ij} - K^{(-)}_{ij} \ .
\end{equation}
The bulk stress tensor integrated over $x$ is equal to the tensor $S_{ij}$, defined by 
\begin{equation}
-8 \pi G_N  S_{ij} = \gamma_{ij} - g_{ij} \gamma_{m}^m \ . \label{Sdef}
\end{equation}

The expression (\ref{Sdef}) for the integrated bulk stress tensor was derived in \cite{Israel} for a 4-dimensional space with no cosmological constant.  While several factors in the derivation change in our current 5-dimensional situation, Eq.~(\ref{Sdef}) remains unchanged.  The cosmological constant term yields a contribution proportional to the integral of $\Lambda$ times the metric integrated over the thickness $d$ of the surface layer.  As the metric is taken to be finite, this term vanishes in the $d\rightarrow 0$ limit.

The trace of $\gamma$ is $O(v^2)$, therefore (\ref{Sdef}) yields the integrated bulk stress tensor
\begin{align}
S_{ty} &= - \frac{v'}{16\pi G_N b^4 r^3} \  \\
S_{yr} &=  - \frac{v'}{16\pi G_N r}\left(1 + b^2 r^2 {\mathfrak g}'_1(br) \right) \ .
\end{align}
These are equal to the integrals over the $x$ direction\footnote{Note that, following Ref.~\cite{Israel}, the measure of this integral must be that of $x$ rescaled to normal coordinates.  Therefore the integral contains an additional factor of $r=\sqrt{g_{xx}}$.} of the stress tensors $T^{(1)}$ of Subsec.~\ref{T1sez} at order $k=1$, in other words, without the ${\mathfrak v}_1$ term that entered into the metric (\ref{metfin}) multiplied by $v\p\p$
\begin{align}
T^{(1)}_{ty} &= - \frac{v''}{16\pi G_N b^4 r^4} \  \\
T^{(1)}_{yr} &=  - \frac{v''}{16\pi G_N r^2}\left(1 + b^2 r^2 {\mathfrak g}'_1(br) \right) \ .
\end{align}
The ${\mathfrak v}_1$ terms arose from the dualization of the interpolating region, which did not satisfy the equations of motion.  It therefore cannot enter into the Israel calculation, which uses only the solutions of the vacuum Einstein equations.  Indeed, the ${\mathfrak v}_1$ terms in (\ref{metfin}) are singular in the limit $d\rightarrow 0$ as $v\p\p$ diverges as $1/d$, and therefore the boundedness of the extrinsic curvature assumed in Israel's derivation fails for the metric interpolation (\ref{metfin}).

Like the stress tensor (\ref{tty}) calculated by interpolating the hydrodynamic flow, the Israel stress tensor does not satisfy the null energy conditions.  Again, to linear order in $v$, one may consider vectors which are null with respect to the static black brane metric.  Therefore, again one may consider the null vectors $w_\pm$ of Eq.~(\ref{w}).  As $T_{ty}$ is, at least for any finite $d$, equal to that of Subsec.~\ref{T1sez} divided by the positive combination $br$, the sign of the inner product (\ref{prod2}) is unchanged.  Therefore the null energy condition is also violated by this stress tensor.  

The main difference between the two stress tensors is then that $T_{yr}$ does not vanish for the Israel tensor.  Remembering that in our Gaussian null coordinates the $r$ direction is the sum of a spatial and temporal piece, the spatial component implies that there is a nonzero stress.  More precisely, while both Israel's thin surface layer and the thick fluid surface layer have a nonvanishing $y$ momentum, the Israel surface layer also has a flux of this $y$ momentum in the radial direction, from the boundary into the horizon of the black brane.  As the black brane is infinite in the $x$ direction, this is not problematic for the time-independence of the solution.

\section{Future directions}

Turbulence often arises as a result of the boundary conditions placed on a fluid.  As a preliminary step towards an understanding of turbulence in gravity, we have proposed two gravitational duals of such boundaries.  Both of these duals involve the addition of a surface layer of matter, with a certain stress tensor.  These proposals are in a sense trivial, as the dynamics of the duals is defined not by any known equations of motion, but by the duality map itself.  It remains to be shown whether such matter can exist.  For example, even if the equations of motion which it obeys can be found, the existence of a UV completion of the matter theory may be fundamentally obstructed as in Ref.~\cite{nima}.  Or the failure of the null energy condition may imply that, whatever the ultraviolet theory may be, the wall simply disintegrates before it has any significant effect on the fluid.  

Of course, an ultraviolet completion is not necessarily a prerequisite for learning something interesting about whatever the gravitational dual to turbulence may be.  After all, no ultraviolet completion of Einstein gravity is used in this correspondence.  The surface layer implies the existence of equations of motion which are distinct from the Einstein vacuum equations and perhaps pathological.  However the interesting part of the fluid, the turbulent part, is not at the wall.  For example, if we consider the motion of a fluid in a pipe, the flow may be turbulent throughout the interior of the pipe.  The ultralocality of the duality map implies that, at a distance greater than $1/T$ from the pipe, the vacuum Einstein equations are still satisfied by the gravity dual.   Thus in a sense the ultralocality decouples the problem of understanding turbulence in gravity from the problem of defining a gravity dual of a boundary.

Besides trying to characterize the gravitational dual of turbulent flow, the other interesting question is to find the gravitational dual of the conditions under which turbulence can occur.  In nonrelativistic, incompressible flows, turbulence is expected when the product of a system's characteristic scale $L$ times the characteristic velocity $v$ of a fluid is much greater than the kinematic viscosity.  In Ref.~\cite{forced}, the authors claim that for the conformal fluids dual to AdS black branes, turbulence is expected when $LTv\gg1$, where $T$ is the temperature of the fluid.  The AdS/hydrodynamics correspondence is expected to be reliable at scales $L$ such that $LT\gg1$.  Therefore since $v<1$, it appears that whenever turbulence is expected, $LT>LTv\gg1$ and so the correspondence can be trusted at least for quantities that vary over a distance $L$.  $(3+1)$-dimensional turbulence is characterized by vortices of various sizes from $L$ down to the dissipation scale \cite{Rich}.  Thus the duality appears to be reliable at least for the largest vortices in a turbulent flow.  The dissipation scale is a function of $L$, $T$ and $v$, and so in principle one may determine whether or not the duality is reliable for vortices all of the way down to this scale and so for the entire flow.

Understanding the gravity duals of turbulent flows, as described above, may yield new insights into the dynamics of black branes in AdS space, perhaps revealing a surprising difference between branes in AdS$_4$ and AdS$_5$, or indicating that generically they come with funnels attached as in Refs.~\cite{MarolfRang}.  The main weakness of this program is the dependence on asymptotically AdS geometry in the duality map of Ref.~\cite{BHMR}.  There was no such restriction in the original correspondence of Ref.~\cite{Damour}, nor in other identifications of black holes and viscous fluids such as the blackfold program of Refs.~\cite{bf1,bf2} and the Wilsonian identification of Ref.~\cite{Strominger}.  An extension of turbulence to asymptotically Minkowski space could relate (3+1)-dimensional fluid dynamics to wealth of studies of asymptotically Minkowski 5d black objects, such as Refs.~\cite{5d}.  More importantly, relaxing the asymptotically AdS condition may mean that fluid mechanics, perhaps in only 2+1 dimensions, has something to teach us about real world gravity.  

\section*{Acknowledgments}

\noindent
We have benefited beyond any reasonable measure from discussions with Juan Maldacena.


\end{document}